\documentclass[12pt,preprint]{aastex}

\newcommand{\tc}{$T_{c}$}
\newcommand{\ttvamplitudelimit}{$1.5$~minutes}
\newcommand{\diffpoint}{124}

\usepackage{color}

\shorttitle{Five new transit epochs of the exoplanet OGLE-TR-111b}
\shortauthors{Hoyer et al.}

\begin{document}


\title{ Five new transit epochs of the exoplanet OGLE-TR-111b\footnote{Based on observations made with ESO Telescopes at the Paranal Observatories under programme ID 278.C-5022}\\ }


\author{S. Hoyer\altaffilmark{1}}
\email{shoyer@das.uchile.cl}

\author{P. Rojo\altaffilmark{1}}
\email{pato@das.uchile.cl}

\author{M. L\'opez-Morales\altaffilmark{2,3}}
\email{mercedes@dtm.ciw.edu}

\author{R.F. D\'iaz\altaffilmark{4,5} }
\email{diaz@iap.fr}

\author{Chambers, J.\altaffilmark{3}}
\email{chambers@dtm.ciw.edu}

\and

\author{Minniti, D. \altaffilmark{6,7}}
\email{dante@astro.puc.cl}


\altaffiltext{1}{Astronomy Department, Universidad de Chile, Casilla 36-D, Santiago de Chile, Chile}

\altaffiltext{2}{Institut de Ciencies de l'Espai (CSIC-IEEC), Campus UAB, Facultat de Ciencies, Torre C5, parall, 2a pl, E-08193 Bellaterra, Barcelona, Spain}

\altaffiltext{3}{Carnegie Institution of Washington, Department of Terrestrial Magnetism, 5241 Broad Branch Rd. NW, Washington, D.C. 20015, USA. }

\altaffiltext{4}{Institut d'Astrophysique de Paris, UMR7095 CNRS, Universit\'e Pierre \& Marie Curie, 98bis boulevard Arago, 75014 Paris,France}
\altaffiltext{5}{Observatoire de Haute-Provence, CNRS/OAMP, 04870 Saint-Michel-l'Observatoire, France}


\altaffiltext{6}{Departamento de Astronom\'ia y Astrof\'isica, Pontificia Universidad Cat\'olica de Chile, Casilla 306, Santiago 22, Chile}

\altaffiltext{7}{Vatican Observatory, V-00120 Vatican City State}


\begin{abstract}

We report five new transit epochs of the extrasolar planet \object{OGLE-TR-111b}, observed in the \textit{v-HIGH} and \textit{Bessell I} bands with the FORS1 and FORS2 at the ESO Very Large Telescope, between April and May 2008.  The new transits have been combined with all previously published transit data for this planet to provide a new Transit Timing Variations (TTVs) analysis of its orbit. 
We find no TTVs with amplitudes larger than \ttvamplitudelimit ~over a 4-year observation time baseline, in agreement with the recent result by \cite{adams10a}. Dynamical simulations fully exclude the presence of additional planets in the system with masses greater than 1.3, 0.4 and 0.5 $M_{\earth}$ at the 3:2, 1:2, 2:1 resonances, respectively. We also place an upper limit of about 30 $M_{\earth}$ on the mass of potential second planets in the region between the 3:2 and 1:2 mean-motion resonances. 

\end{abstract}


\section{Introduction}

The method of Transit Timing Variations (TTVs) has been proposed by several recent theoretical works as of great potential to detect additional exoplanets in transiting exoplanetary systems \citep{miralda02,agol05,holman05}, and even exomoons \citep{sartoretti99,kipping09a}. In a system with additional planets, the predicted central time of a transit can change periodically by a significant amount of time if a second planet, the perturber, is in mean motion resonance with the transiting one. In the case of exomoons, TTVs of up to few minutes can be produced by moons larger than 1 $M_{\earth}$ depending on the physical parameters of the orbital system. Variations of other transit parameters, such as the depth or duration of the transits, have been predicted to also indicate the presence of additional planets in those systems \citep{miralda02}. Another application of the TTV technique is the detection of long-term orbital period secular variations produced by tidal interactions between the star and the planet, star oblateness and general relativity. Those effects  are predicted to introduce changes in the orbital periods of the transiting planets with amplitudes of the order of 0.3-10 ms/yr \citep{miralda02,HandG07,jordan08} and would be detectable in 10-20 years timescales given the precision of the current techniques.

These theoretical predictions have prompted the work of several observational groups, who in the past few years have started to monitor various transiting exoplanets from the ground. More recently, dedicated transit search space missions like CoRoT \citep{barge08} and Kepler \citep{borucki10} have also started to collect high duty cycle and long-term monitoring data which allow TTV studies of the new transiting planets they find \citep[e.g.][]{bean09}.   Most of the findings from observational efforts are still preliminary, but include the potential detection of TTVs with amplitudes of up to 3 min for the hot Jupiter WASP-3b, which could be explained by the presence of a $\sim$ 15 $M_{\earth}$ planet near the outer 2:1 mean motion resonance with WASP-3b \citep{Maci10}, and the also preliminary detection of a  $-60\pm15$ ms/yr orbital period decay of the hot Jupiter OGLE-TR-113b by \cite{adams10b}. 
  Other preliminary results include the hints of transit duration and orbital inclination variations in the hot Neptune \object{Gliese 436b} system reported by \cite{coug08} and the shift of about 3 seconds in the orbital period of XO-2b reported by \cite{fernandez09}.   Several other system have been monitored without showing any clear evidence of TTVs, e.g. \cite{SteffenAgol05} found no evidence of a second planet using timing data of eleven transits of \object{TrES-1b}, and \cite{AgolSteffen07} ruled out Earth-mass planets in low order resonances in \object{HD 209458b} system.  In a later work, \citealp{miller08a,miller08b} found no TTVs with amplitudes larger than 45 seconds on the transits of  \object{HD 209458b} and \object{HD 189733b} based on \textit{MOST} data, and \cite{winn09} found no variations from a constant orbital period for \object{WASP-4b}, after combining two precise mid-transit timing measurements with another five measurements for this planet reported by \cite{gillon09}.  Recently, \cite{holman10} announced the first unquestionable evidence of TTVs in the double transiting planetary system \object{Kepler-9}.

OGLE-TR-111b was the first hot Jupiter for which tentatively significant TTVs were reported by \cite{M07}, hereafter M07. The first two precision transit timing data points for this planet were published by \cite{W07}, W07. Shortly after, M07 published a third data point which deviated by about 5 minutes from the predicted W07 transit ephemerides. In a subsequent paper, \cite{D08} (D08) reported two new consecutive transits and combined their new data with the other three epochs to find TTVs with amplitudes of up to 2.5 minutes, which the authors suggested could be explained by the presence of a 1.0 $M_{\earth}$ perturbing planet in an outer eccentric orbit to OGLE-TR-111b. The most recent TTVs analysis publication on this planet \citep[][A10]{adams10a} reports six new transit observations and no TTVs for this planet with amplitudes larger than $71 \pm 67$~seconds.

In this new work we present five additional transits of OGLE-TR-111b observed between April and May 2008, and we perform a new homogeneous timing analysis of all available 16 epochs to further study the presence or absence of additional planets in this system. In section \ref{observaciones} we present the new observations and the data reduction. Section \ref{fiteo} describes the modeling of the light curves. In sections \ref{ttv} and \ref{otherparameters} we describe the timing and possible parameters evolution.  In  \ref{masslimits} we discuss the mass limits for a \textit{unseen} perturber and finally in section \ref{conclusiones} we present our conclusions.

\section{Observations and data reduction \label{observaciones}}

\begin{figure}
\plotone{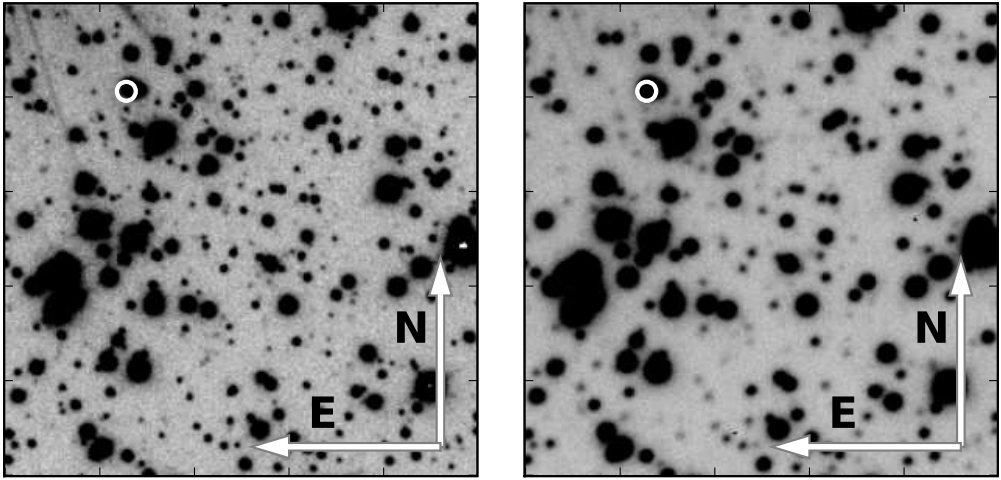}
\caption{Portion of $0.5\arcmin \times 0.5\arcmin$ images of the night 2008-05-03 observed with FORS2 at VLT.    The best image (FWHM$\sim 0.34\arcsec$) of the night is shown at the left panel and the worst image (FWHM $\sim 0.55\arcsec$ ) is shown at the right panel. The location of OGLE-TR-111 is marked by the circle.  Due to the seeing and pixel scale our target appears blended with a nearby star.  Also one of the diffraction spikes of a very bright star is visible at the upper left corner of the best seeing image.  Ocassionally, these spikes reached the location of the target contaminating the photometry. \label{FoV}}
\end{figure}

\begin{figure}
\plotone{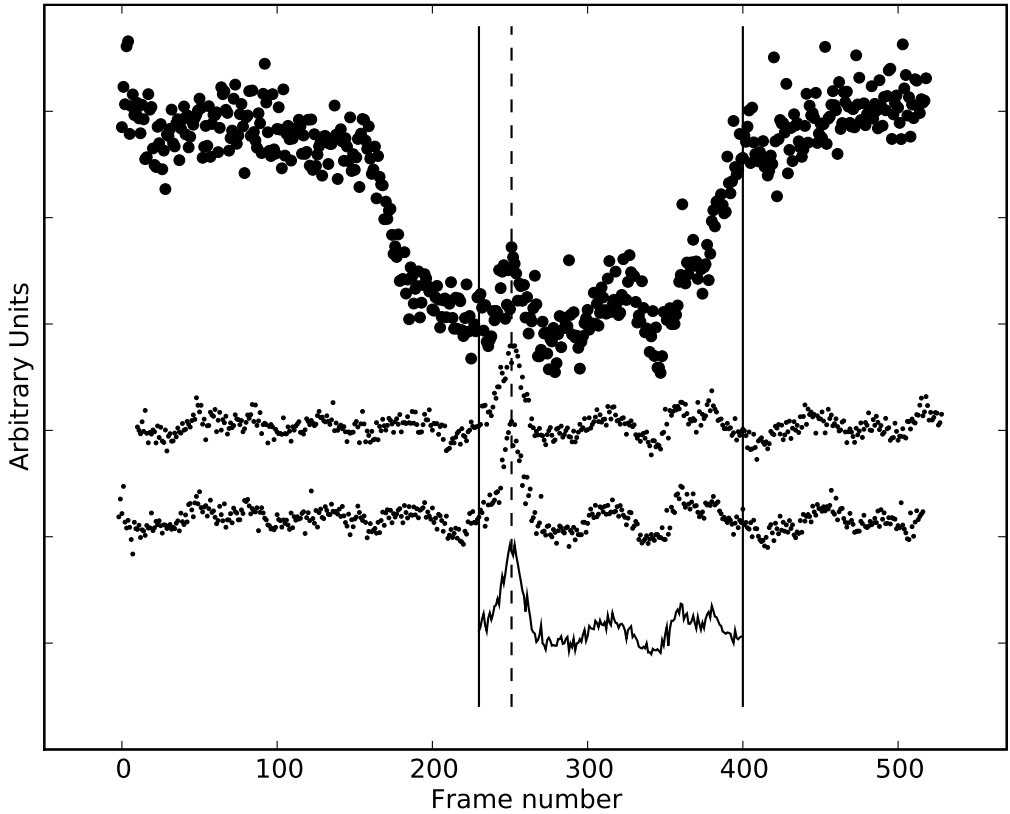}
\caption{Light curve of OGLE-TR-111 (top) and two comparison stars (middle) of the night 2008-05-03 observed with FORS2 at VLT. The contamination produced by the diffraction spikes of a very bright star is evident in the region enclosed by the solid vertical lines.   The average flux of the comparison stars (bottom) after align the peaks of the bumps (vertical dashed line) was substracted from the light curve of the transit to remove the contamination. This procedure was only applied to this transit.  \label{spikes}}
\end{figure}

Between April and May 2008, we observed five transits of the exoplanet \object{OGLE-TR-111b} with FORS1 and FORS2 \citep{appen98} at the ESO Very Large Telescope. The first four transits were fully covered in phase. The fifth transit was only partially covered because the target reached the telescope's airmass limit, and is only complete between phases 0.97 and 1.01, which includes the out-of transit baseline before the transit, the ingress and most of the bottom of the transit.  The UT date of the mid-time of the transit, instrument, filter band, exposure time, airmass range and number of frames of each observation are summarized in Table~\ref{tabla-obs}. FORS1 and FORS2   are visual focal-reducer imagers composed of two $2048\times4096$ E2V/MIT CCD detectors mosaics with a pixel scale of 0.126 arcsec pixel$^{-1}$ each (high resolution mode). The field of view (FoV) of each camera is therefore $4.25\times4.25$ arcminutes, large enough to include OGLE-TR-111 and several comparison stars.  FORS1 and FORS2 have the same wavelength coverage (3000-11000 \AA) but FORS1 is optimized for blue wavelengths ($< 5000$ \AA ) while FORS2 is for the red ($> 6000$ \AA). Our first transit was observed with FORS1 using a \textit{v-HIGH} filter ($\lambda_{eff}=557$~nm) while the rest of the transits were observed with FORS2 using a \textit{Bessell I} filter ($\lambda_{eff}=768$~nm).  A very close star appears partially blended with the target given the resolution of the instrument and the typical seeing conditions during the observations (Figure \ref{FoV}).  The center of the field was selected so that \object{OGLE-TR-111} and several good comparisons stars would fall on a single detector, while also locating a bright nearby star out of the field. However, diffraction spikes from that bright star moved across the field of view, occasionally reaching the location of the target, as illustrated in Figure \ref{FoV}. Our subsequent analysis revealed additional background noise in some of the images due to this effect. This problem was most evident in the 2008-05-03 light curve (Figure \ref{spikes}) where pronounced bumps were visible in the bottom of the light curve of this night. Along the images obtained during this night, the diffraction spikes rotated trough the FoV reaching the comparison stars at different phases of the transit.   The bumps also appeared in the light curves of these comparison stars.   We use the peak of the larger bump in the light curve of the target to match in phase the bumps of all the other light curves (see Figure \ref{spikes}).  We average the light curves of the comparison stars in a small region where the bumps were more evident (between the horizontal lines in Figure \ref{spikes}) and finally in order to remove the bumps, we substracted this average to the light curve of the transit.  In the other nights it was not possible to identify clearly the bumps produced by the diffraction spikes, and therefore we could not reproduce the process mentioned before but we attributed some of the noise in the lights curve to this effect. 

We worked with the processed data provided by the \textit{VLT} pipeline which performs the bias and flatfield corrections. The times at the start of the exposure are recorded in the images headers, in particular we used the value of the \textit{Modified Julian Day} keyword of each image and transformed it to \textit{Barycentric Julian Day} (see section \ref{ttv} for details). 

  Our photometric analysis was done with the Difference Image Analysis Package (DIAPL) written by \cite{W2000} and recently modified by W. Pych\footnote{The package is available at http://users.camk.edu.pl/pych/DIAPL/index.html}. The package is an implementation of the method developed by \citet{alard98}, and it is optimized to work with very crowded fields and/or blended stars, as is the case of \object{OGLE-TR-111} (Figure \ref{FoV}).   DIAPL models PSF variations along the X-Y coordinates, scales and subtracts the flux of the stars of a template image for each frame, among other calculations.  One of the disadvantages of this code is the time required to complete the process, specially since it is not possible to verify the suitability of the input parameters and the quality of the products until the last steps (subtraction and/or photometry).  In our case, due the large size of the frames and the large number of stars in the FoV, one iteration required up to 20 minutes.  To eliminate a few nearby saturated stars in the field that hindered the photometry and to reduce considerably the processing time we worked on  $ \sim 500 \times 500 $ pixel subframes.  Working with these subframes, DIAPL estimations of the PSF, background and flux levels are more representative of the vicinity of our target in each frame and in the reference image; noise as well as obvious systematics in the final photometry are considerably reduced.  Reference frames for each night were constructed by combining the 20-27 best images (in terms of seeing and signal-to-noise) depending on the night's conditions.  We used aperture radii between 4 to 10 pixels (in DIAPL's task \textit{phot.bash}) to perform the relative photometry of the target and comparison stars (8 to 15 stars) on each subtracted frame.

To obtain an absolute normalization, we performed aperture photometry (DAOPHOT / ALLSTARS, \cite{stetson87}) in the reference frame of each night using a curve of growth analysis to select the ideal aperture radii. 

The resultant light curve contains some remaining systematic variations, which we have modeled using linear regression fits of the out-of-transit data points against the airmass, average FWHM of the point spread function and/or background level around our target of each frame.  Doing this we were able to achieve an \textit{RMS} precision of $\sim 0.0013-0.0027 ~mag$ in the light curves (almost reaching the Poisson noise level in the best nights or doubling it in the worst).   

\section{Modeling Light Curves \label{fiteo}}

We fitted our five new light curves together with all the light curves previously published by W07, D08 and A10, using JKTEBOP\footnote{http://www.astro.keele.ac.uk/~jkt/codes/jktebop.html} \citep{south04}. The fit also includes the light curve by \citet[P10 hereafter]{pawel10}, obtained from the reanalysis of the VIMOS data published by M07. P10 found a problem with the times reported by M07, which results in a mid-transit epoch time difference of $\sim$ 5 minutes. We only consider the P10 analysis of this transit from this point onwards.

Among the parameters fitted by JKTEBOP for each light curve are: the planet to star radii ratio ($k$), the inclination ($i$) and eccentricity ($e$) of the orbit, the out-of-transit baseline flux ($F_b$), the mid-time of transit ($T_c$), the quadratic limb darkening coefficients ($u_1$ and $u_2$), and the sum of the fractional radii, $r_p + r_s$. The terms $r_p$ and $r_s$ are defined as $r_p = R_p / a$ and  $r_s = R_s / a$, where $R_p$ and $R_s$ are the absolute stellar and planetary radii, and $a$ is the orbital semi-major axis. In the case of the limb darkening coefficients, we fixed $u_2$ to the values given by \cite{claret00} and \cite{claret04} for each observation's filter (for the \textit{v-HIGH} filter we use the $v$ coefficient) and only left $u_{1x}$ as free parameter during the fitting process described below, where $x$ denotes the filter band.  We performed the same fitting method using a linear limb-darkening law obtaining basically the same final $\chi^{2}_{red}$ for each transit, which reveals that the photometric precision of the light curves is not sufficient to distinguish between limb darkening laws. Also, to minimize potential degeneracies between parameters, we fixed the eccentricity and the longitude of the periastron of the orbit to zero, and the planet to star mass ratio to $m_p/m_s = 0.00061$, adopting the mass values derived by \cite{santos06}.

\begin{figure}
\plotone{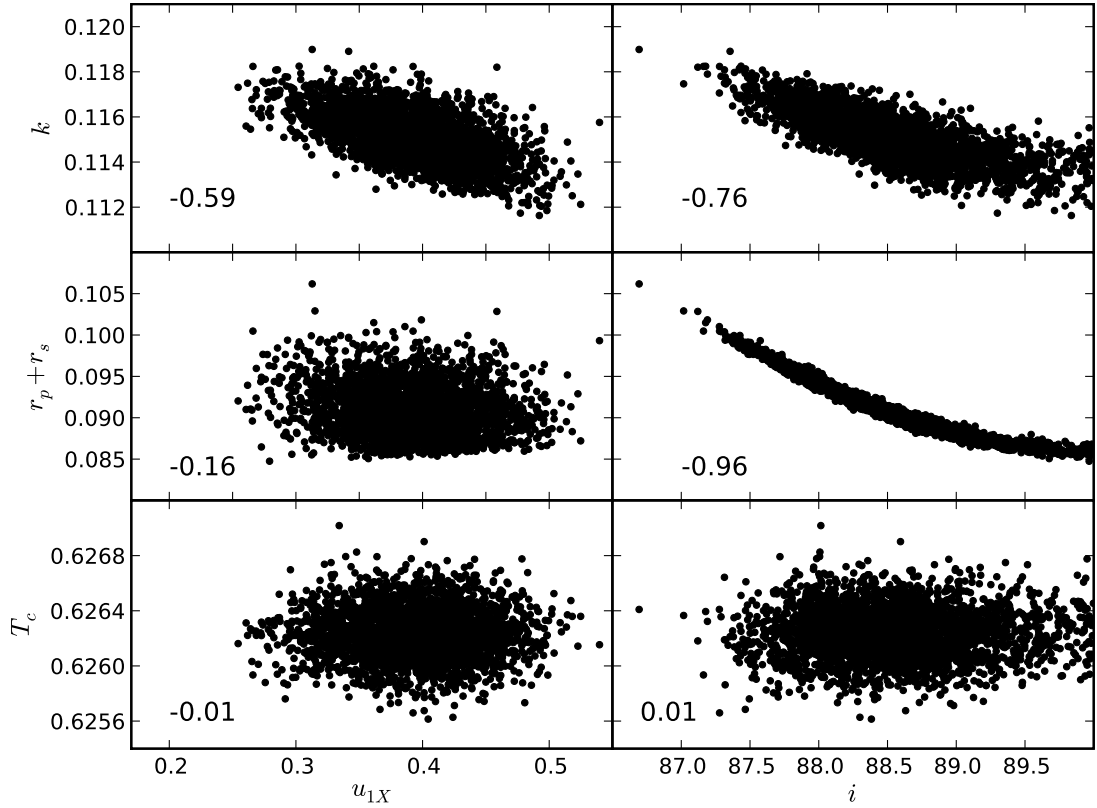}
\caption{3000 results of 10000 Monte Carlo iterations (using JTKEBOP with the data of the night 2008-05-12) which show the correlation between the parameters fitted from the light curves: $k$, $r_{p}+r_{s}$, \tc, $i$ and the linear coefficient $u_{1X}$ of a quadratic limb darkening law, when all are left variable.  The correlation coefficients between each variable are show in the bottom left corner of each box. \label{correlacion}}
\end{figure}

JKTEBOP also allows for an statistical determination of the error of each parameter, as well as an analysis of the impact of systematics in the light curves, via Monte Carlo simulations. We ran $10^4$ Monte Carlo simulations adding random simulated gaussian noise to the input parameters to estimate the uncertainty of each parameter while also testing for potential correlations. When all the parameters described above are left free, there are clear correlations between $k$, $i$, and $u_{1x}$, and also between $r_p + r_s$ and $i$, as illustrated in Figure \ref{correlacion}. No significant correlation was observed between any of the other parameters.

\begin{figure}
\plotone{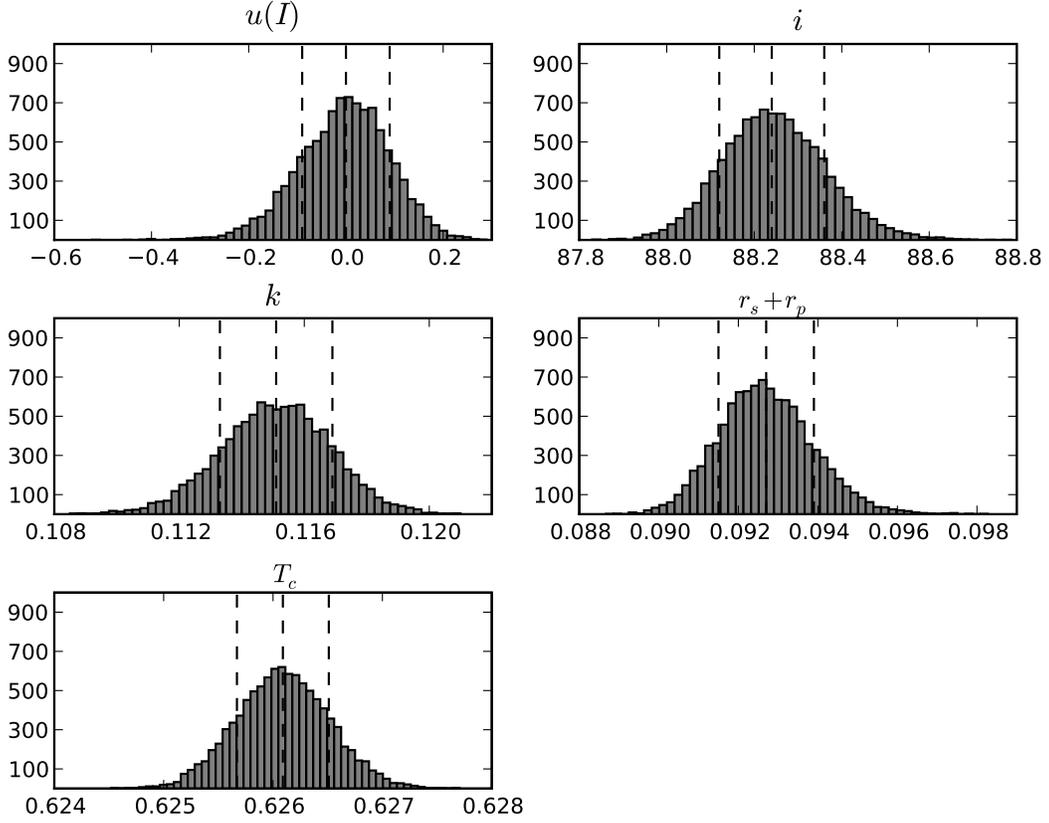}
\caption{Histograms of the 10000 Monte Carlo iterations (using JTKEBOP with the data of the night 2008-05-12) for the fitted parameters on each of the : $k$, $r_{p}+r_{s}$, \tc, $i$ and the linear coefficient $u_{1x}$ of a quadratic limb darkening law obtained after the steps I, II and III described in Section \ref{fiteo}.  The dashed lines show the fitted value and the $\pm 1\sigma$ errors which were compared with the red noise estimation using the prayer-bead method.  The same analysis was performed on each of the 16 transits. The final value of $k$, $r_{p}+r_{s}$, $i$, $u_{1i'}$ and $u_{1I}$ corresponds to the weighted average of these results, except for $u_{1v-HIGH}$ and $u_{1V}$ where we adopted the results of JKTEBOP (see text). \label{histograms}}
\end{figure}

These correlations can be minimized by fitting the parameters in three steps, also running $10^4$ Monte Carlo simulations on each step.  First, we fixed the inclination of the orbit to the value obtained by A10, $i = 88.3\degr$ together with the corresponding value of $u_{1x}$ from the Claret tables and fitted $k$ and $r_p + r_s$ for each light curve. Next, we fixed  $k$ and $r_p + r_s$ to the weighted average of the individual fits obtained in step I, and left $i$ and $u_{1x}$ as a free parameters. Finally, we adopted the weighted average of the resulting inclinations and the corresponding $u_{1x}$ for each filter (because we have only one transit observed with \textit{v-HIGH} and one with V we adopted directly the results of JKTEBOP for its $u_{1x}$), and fitted only for $T_c$. In Table \ref{pasos} we summarized each step of the fitting process.  An example of the histograms of the distribution of values for each parameter for the night 2008-05-12 are represented in Figure \ref{histograms}.\notetoeditor{The figure 4 can be converted to one column plot if necessary } The adopted values of each parameter for the individual light curves are summarize in Table \ref{tabla-par}. The average values for the system based on all light curves are summarized in Table \ref{tabla-final}.
To test the consistency of the Monte Carlo error estimates, we compared it to the results of the prayer-bead method \citep{Bouchy05}. While the Monte Carlo method gives an idea of the white noise in the data, any level of red noise (time correlated noise, ) is best characterized by the prayer-bead method. The errors obtained by the prayer-bead method were, in general, larger than the Monte Carlo errors, showing that red noise is the dominant factor in the light curves. In Table \ref{tabla-par} we show the ratio $R$ of the errors estimated by the Prayed-Bead and the Monte Carlo method.  We adopted as 1$\sigma$ errors of the parameters reported in Table \ref{tabla-par}, the larger values between these two error estimations. 


The Levenberg-Marquardt Monte Carlo (LMMC) fitting method implemented by JKTEBOP, can present some disadvantages with respect to the Markov Chain Monte Carlo (MCMC), method used by several other recent TTVs studies.  For example, LMMC can be trapped in a local mimima or/and  can underestimate the errors of the fitted parameters \citep[e.g.][]{fisher09,driscoll06}. Despite this, our results in Table \ref{tabla-par} for the analysis of previously published light curves are fully consistent with the parameter fit values reported in Table 5 of A10 using MCMC. LMMC and MCMC are expected to yield similar results in well-behaved parameter space, i.e. with no multiple minima, as seem to be the case of our dataset. 
Even though, it is possible that the errors of LMMC best-fit parameters (provided by the parameter distribution of the Monte Carlo iterations) can be under estimated, but our adopted parameter errors are dominated/scaled by the red noise contribution (see the $R$ values in Table \ref{tabla-par} ), indeed our errors are very conservative in comparison with A10 estimations.

\section{Timing Analysis \label{ttv}}

The mid-time for each transit derived in the previous section and listed in Table 2 was converted to Barycentric Julian Days, expressed in terrestrial time, i.e. BJD(TT), following the standard timing reference system recommendations by \cite{eastman10}. Since our analysis includes data from different instruments/telescopes, spanning over four years, and reduced by different groups we checke carefully for any possible systematics. As example, A10 has already pointed out how previously  published results on \object{OGLE-TR-111b} had not corrected the reported times for leap seconds (UTC to TAI) nor for the 31.184 second conversion between TAI and TT, where TAI is defined as the $International$ $Atomic$ $Time$. We have applied the same corrections to all the literature light curves used in our analysis.

As an additional check to our reduction, light curve fitting and timing correction procedure, we compared our final BJD(TT) times to those published by A10. Particularly valuable for this test is the 2008-05-12 transit epoch, which was independently observed by us and  A10. As illustrated in Figure \ref{comparacion}, in spite of adopting completely different approaches to fit the light curves, all our derived transit mid-times agree well with the values obtained by A10.  When comparing the \tc ~of our 2008-05-12 transit with A10's, both times agree within the $1\sigma$ error of our observation, although our mid-transit time occurs \diffpoint ~seconds earlier.   Four of the five transits we measure also produce mid-transit times on average 100 seconds earlier than the ephemerides predict (see Figure \ref{comparacion}), although the differences are not statistically very significant.  However, we have decided to further investigate the potential source of this discrepancy.   First, we confirmed with the VLT staff that the times recorded in the image headers are the UTC times at the beginning of each exposure of our data (including all leap seconds). We further tested for any potential systematics between datasets by binning our data to 60 seconds, and removing the last $\sim$ 50 points in our light curve to make it match the light curve sampling and phase coverage of A10. The result was only a 10 second time shift in the resulting $T_{c}$ compared with the previous fit value. The two transits were observed in different filters, but we find no correlations between the limb darkening coefficients and \tc ~(see Figure \ref{correlacion}) that could account for this mid-time discrepancy. We attributed this difference to the red noise in the FORS light curve, possibly due to weather (the $R$ value of \tc~of this transit in Table \ref{tabla-par} is almost twice the value of A10 light curve).


Figure \ref{o-c}a shows the updated $Observed$ $minus$ $Calculated$ (O-C) diagram with all the final BJD(TT) mid-time values calculated from the literature light curves and our five new transits. The data show a linear trend which can be attributed to the accumulation of timing uncertainties with respect to the adopted transit ephemerides (D08). After removing that linear trend (Figure \ref{o-c}b), the data are consistent with a constant period ephemeris equation of the form:

\begin{equation}
T_{c}=2454092.80691(25)[BJD] + 4.0144477(16) \times N, 
\end{equation}


where \tc ~is the central time of a transit in $N$ epochs since the reference time $T_{0}$. This fit has a reduced $\chi^2$ of 1.8, so the errors in \tc ~have been rescaled by a factor of $\sqrt{\chi^2} = 1.18$ to make them consistent with $\chi^2 = 1$ (see errorbars in Figure \ref{o-c}b), The new period is fully consistent with the one obtained by A10.

\section{Analysis of additional parameters of the light curves \label{otherparameters}}

We tested for possible variations of the physical parameters of the \object{OGLE-TR-111} system using the values of the transit
duration, $T_{14}$, the inclination, the planet to star radius ratio, and the sum of the fractional radii derived for each transit with JKTEBOP.  The value of each of those parameters over time is represented in Figure \ref{ki}.

There is no evidence of trends for any of the inspected parameters.  However, it is noticeable that the results from the light curves observed by different groups appear clustered around the same values.  We attribute that clustering to systematics introduced by the way the photometry is performed (i.e. aperture or PSF photometry, differential image analysis, and so on), and probably also by the way the light curve systematics are treated by the different groups.   For example, D08 already pointed out that the depths of their transits were smaller than the average and that this was due to a reduction artifact of the differential image subtraction techniques previously  noticed by \cite{gillon07}. Although we used DIAPL instead of aperture photometry, our results are consistent with those of A10. Notice, however, that systematics in the transit depths have no effect in the determination of the transit midtimes.

\section{Limits to additional planets \label{masslimits}}

Based on the timing constrain of our $O-C$ diagram, i.e. no TTV variations with amplitudes larger that \ttvamplitudelimit ~over a three year period, we run dynamical simulations to place limits on the mass and the semi-major axis of a possible orbital companion of the transiting planet using the \textit{mercury} code \citep{chambers99}.  The first step was to explore stable orbital regions by assuming  a massless point particle over a range of initial semi-major axes, and fixing all the other input variables to the known physical parameters of the system.  The orbital evolution of the massless particle was integrated over $10^{6}~days$.  This test yields a strip of unstable orbits between  $0.034-0.056 ~AU$, where encounters between the test particle and \object{OGLE-TR-111b} would occur.  For all the other orbits we calculate TTVs of the transiting body with the hypothetical coplanar perturber using a wide range of masses ($0.1 ~M_{\earth} \leq M_{per} \leq 5000 ~M_{earth}$), variable density (from Earth to Jupiter density depending on the mass) and semi-major axes ($0.02 ~AU \leq a \leq 0.13 ~AU$ in steps of $0.005 ~AU$) with $\sim~4500$ simulations over 7 years.  All the initial relative angles were fixed to zero. Near resonances, the steps in the variables were reduced to increase precision.  Only the last 5 years were used for the timing analysis, since that is about the same time span covered by the observations, and also to minimize any effects introduced by the choice of initial parameters.   Then we calculated the central time of each transit during all the simulations.
Similarly to what was done in section \ref{ttv}, we did a linear fit of these central times and we defined the TTV of each simulation as the standard deviation of the central times with respect to this linear fit.   We checked that the final period of the transiting planet did not change by more than 3$\sigma$ from its initial value due the gravitational interaction.        
With this method we were able to make a $mass$ vs $a$ diagram (Figure \ref{mvsa}-A) where the solid line represents TTVs of 1.5 minutes. For comparison we also plot the mass limits of TTVs of 0.5 and 5 minutes (dotted and dash-point lines). The dashed line corresponds to the detectability limit placed by radial velocity observations \citep{santos06}. Our mass constrains are upper limits since for perturbers with an orbital eccentricity different from zero, the mass necessary to produce TTVs of the same amplitude will be lower. We confirm this by performing a set of simulations with $e=0.3$  and using as input parameters the values ($M_{per}$, $a$) which produced TTVs $\sim 1.5 ~min$ in the case $e=0$ (see Figure \ref{mvsa}-B).   With this configuration the unstable region becomes wider due to encounters between the orbital bodies and the TTVs produced for a given mass were larger than in the case of $e=0$. Same results were obtained setting the initial values of the longitude of the periastron different from zero ($90\degr$,~$180\degr$ and $270\degr$) and $e=0.3$. Combining TTV \textit{RMS} and radial velocities we can rule out the presence of a perturber body with mass greater than 1.3, 0.4 and 0.5 $M_{earth}$ at the 3:2, 1:2, 2:1 resonances with OGLE-TR-111b and lower the upper limit in the region exterior to the planet until $\sim  0.08$~AU to companions of less than $\sim30~M_{\earth}$.

\section{Conclusions \label{conclusiones}}

We present 5 new transit light curves of \object{OGLE-TR-111b}.  We homogeneously model all available light curves in the literature and search for any variation in the timing of the transits.  With our updated ephemeris equation we find no TTVs with amplitudes larger than 1.5 minutes and therefore we rule out the presence of a companion in the 2:1, 3:2 and 1:3 orbital resonances.  If the system has an additional orbiting body, its mass has to be lower than 30 $M_{\earth}$ if is located  between 3:2 and 1:2 resonances.  The mass limits we place with our dynamical simulations based in the TTV data are lower than the limits obtained with radial velocities alone.  We search for any trend in the duration, depth of the transit and inclination of the orbit but we do not see any clear evidence of variation with statistical significance. We point out that systematics of no evident source in the observations, reduction and/or analisys processes can induce differences in the values of the parameters obtained from the light curves and therefore a monitoring of transiting exoplanets carry out by the same group can contribute to reduce these differences. 

\section{Acknowledgements}

S.H. and P.R. acknowledgements support from Basal PFB06, Fondap \#15010003, and Fondecyt \#11080271. Additionally, S.H, recieved support from  GEMINI-CONICYT FUND \#32070020, ALMA-CONICYT FUND \#31090030.  M.L.M. acknowledgements support from NASA through Hubble Fellowship grant HF-01210.01-A/HF-51233.01 awarded by the STScI, which is operated by the AURA, Inc. for NASA, under contract NAS5-26555.  We thanks to P. Pietrukowicz for the help in the use of DIAPL and to E. Adams for helpful discussions about her OGLE-TR-111b TTVs analysis.

\clearpage

\begin{deluxetable}{lclcll}
\tabletypesize{\scriptsize}
\tablecaption{Observational information of each night.\label{tabla-obs}}
\tablewidth{0pt}
\tablehead{
\colhead{Transit Date} & 
\colhead{Instrument} &
\colhead{Filter} &
\colhead{Integration Time [s]} &
\colhead{airmass range} &
\colhead{\# images\tablenotemark{c}} 
}
\startdata
2008-04-26 &  FORS1 & v-HIGH    &   30  & 1.25 - 1.43 & 323 \\
2008-04-30 &  FORS2 & Bessell I &   12  & 1.25 - 1.59 & 488  \\
2008-05-04 &  FORS2 & Bessell I &   12  & 1.25 - 1.75 & 522(2)   \\
2008-05-12\tablenotemark{a} &  FORS2 & Bessell I &   4    & 1.25 - 2.18 & 601(94) \\
2008-05-20\tablenotemark{b} &  FORS2 & Bessell I &   8    & 1.29 - 2.07 & 373 \\
\enddata
\tablenotetext{a}{This transit was also observed by \cite{adams10a}.}
\tablenotetext{b}{This transit has a incomplete phase coverage.}
\tablenotetext{c}{The number of images descarted in the analisys is shown in parenthesis.}

\end{deluxetable}

\clearpage

\begin{deluxetable}{cccc}
\tabletypesize{\scriptsize}
\tablecaption{Summary of each step of the fitting process. See section \ref{fiteo} for details.\label{pasos}}
\tablewidth{0pt}
\tablehead{
\colhead{} & 
\multicolumn{2}{c}{Parameters} &
\colhead{} \\
\colhead{Step} & 
\colhead{Free} & 
\colhead{Fixed} & 
\colhead{Reference\tablenotemark{a}} 
}
\tablecolumns{4}
\startdata
I       &    $k$, $r_p+r_s$, \tc &  $i$      & A10       \\
        &                        & $u_{1x}$  & Claret   \\  
II      &   $i$, $u_{1x}$, \tc   &  $k$, $r_p+r_s$       &  Step I   \\
III     &    \tc                 &  $k$, $r_p+r_s$   & Step I   \\
        &                        &  $i$, $u_{1x}$   & Step II  \\

\enddata
\tablenotetext{a}{This column shows the origin of the adopted value of the fixed parameter.}

\end{deluxetable}

\clearpage

\begin{deluxetable}{lccccccccccc}
\tabletypesize{\scriptsize}
\rotate
\tablewidth{0pt}
\tablecaption{Adjusted parameters for each transit using JKTEBOP code.\label{tabla-par}}
\tablehead{
\colhead{Transit date\tablenotemark{a}} &
\colhead{$k$} & 
\colhead{$R$\tablenotemark{b}} & 
\colhead{$r_{p}+r_{s}$}   &
\colhead{$R$} & 
\colhead{$i ~[\degr]$} & 
\colhead{$R$} & 
\colhead{$u_{1x}$ }& 
\colhead{$R$} & 
\colhead{$T_{c}-2450000$ (BJD)} &
\colhead{$R$} & 
\colhead{$\chi^{2}_{red}$ } 
}

\startdata
2008-04-26(122)  &$0.1203(23)$&0.6  & $0.0926(17)$&0.8  & $88.35(14)$&0.4&$ 0.81(05)$&0.5 & $ 4582.56853(48)$&0.5 & $0.17$ \\
2008-04-30(123)  &$0.1168(15)$&1.1  & $0.0929(13)$&1.3   & $88.22(12)$&1.3&$ 0.20(08)$&1.3 & $ 4586.58288(72)$&2.2 & $0.57$ \\
2008-05-04(124)  &$0.1169(21)$&1.6  & $0.0947(19)$&2.0   & $88.49(15)$&1.5&$ 0.22(05)$&0.9 & $ 4590.59853(94)$&3.0 & $0.69$ \\
2008-05-12(126)  &$0.1151(29)$&1.7  & $0.0927(19)$&1.6   & $88.24(27)$&2.3&$ 0.00(28)$&3.0 & $ 4598.6261(18)$&4.2 & $0.66$ \\    
2008-05-20(128)  &$0.1265(24)$&1.4  & $0.109(17)$&1.2   & $89.37(99)$&0.8&$ 0.13(13)$&2.0 & $ 4606.65482(89)$&1.9& $1.34$ \\  
\hline
2005-04-09(-155)  & $0.1179(87)$&1.1& $0.0990(69)$&1.2    &$88.89(53)$&0.7  & $0.89(08)$&1.5 & $3470.5684(11)$&1.5 & $1.08$\\
2006-02-21(-76)   & $0.1266(15)$ &0.8& $0.0934(08)$&0.8 &$88.70(36)$&0.4  & $0.45(05)$&0.9& $3787.70928(56)$&1.9 & $1.15$\\
2006-03-05(-73)   & $ 0.1246(12)$&0.8& $0.0934(08)$&1.2  &$88.78(16)$&1.2  & $0.61(04)$&1.1 & $3799.75213(81)$&2.9 & $1.08$\\
2006-12-19(-1)    & $ 0.1122(19)$& 0.8& $0.0899(12)$&1.0  &$87.85(17)$&1.7& $0.03(25)$&2.4 & $4088.7922(19)$&4.9& $1.3$\\
2006-12-23(0)     & $ 0.1139(26)$& 1.5& $0.0910(11)$&1.0  &$87.97(21) $&2.1  &  $0.06(22)$&2.3 & $4092.8056(15)$&3.7&$1.14$\\
\hline
2008-04-18(120)   & $ 0.1245(31) $ &2.1&$0.098(16)$&1.4  & $ 87.92(81)$&17.3 & $0.26(12)$&1.6 & $4574.54272(83)$&2.3& $0.99$ \\
2008-04-22(121)   & $ 0.1190(21)$&0.8&$0.0944(16)$&1.0  & $ 88.54(17)$&0.9 & $0.40(07)$&0.8& $4578.55497(43)$&1.0& $0.96$\\
2008-05-12(126)   & $ 0.1183(13)$&1.2& $0.0915(09)$&1.3   & $ 88.14(10)$&1.4 &$0.26(07)$&1.5 & $4598.62754(47)$&1.9& $1.02$\\
2008-05-16(127)   & $ 0.1214(16)$& 1.5&$0.0922(14)$&1.8     & $ 88.38(13)$&1.5 & $0.35(07)$&1.7 & $4602.64167(47)$&1.7& $0.99$\\
2009-02-17(196)   & $ 0.1187(15)$&1.7& $0.0926(15)$&2.5     & $ 88.27(17)$&3.0 & $0.27(09)$&2.7& $4879.63863(55)$&2.7& $1.01$\\
2009-03-13(202)   & $ 0.1236(09)$&1.0& $ 0.0893(06)$&0.9     &$ 88.13(13)$&1.5 & $0.28(08)$&1.4 & $4903.72566(26)$&1.0& $1.01$ \\

\enddata
\tablecomments{The values shown in parenthesis correspond to the errors in the last digits.}
\tablenotetext{a}{The Epoch number is shown in parenthesis.}
\tablenotetext{b}{$R$: Prayer-Bead and Monte Carlo errors ratio. See text for details. } 
%
\end{deluxetable}

\clearpage

\begin{deluxetable}{lcc}
\tabletypesize{\scriptsize}
\tablecaption{Final Values \label{tabla-final}}
\tablewidth{0pt}
\tablehead{
\colhead{Parameter} & 
\colhead{Adopted Value} &
\colhead{Error} 
}
\startdata
$k$ & 0.1213     &  $\pm 0.0004  $    \\
$r_{p}+r_{s}$ & 0.0917    &  $\pm 0.0003 $      \\
$i ~[\degr]$ & 88.29     &  $\pm 0.04    $      \\
Period [days] &   4.0144477 & $\pm 0.0000019$        \\
$T_{o} (BJD)$ & 2454092.80691 &  $\pm 0.00030$       \\
$u_{1v-HIGH}$ & $ 0.81$ & $  \pm 0.05$ \\
$u_{1I}$ & $ 0.426$ &  $\pm 0.024$ \\
$u_{1i'}$ & $ 0.313$ &  $\pm 0.033$ \\
$u_{1V}$ & $ 0.89$ & $  \pm 0.08$ \\

\enddata
\end{deluxetable}

\clearpage

\clearpage

\clearpage

\clearpage

\clearpage

\begin{figure}
\epsscale{.75} 
\plotone{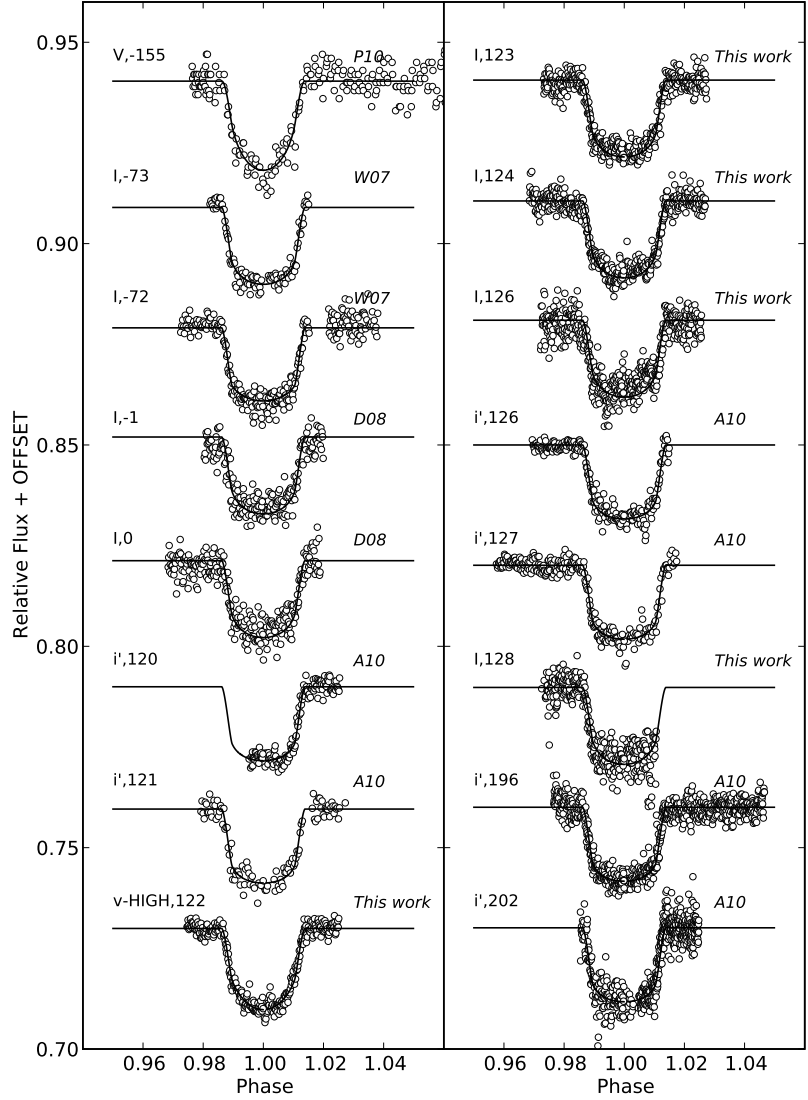}
\caption{ Light curves of all transits of OGLE-TR-111b. The solid lines show our best model fits produced by JKTEBOP. The filter, epoch number and author of the light curve is also indicated. 
\label{lcall}}
\end{figure}

\clearpage

\begin{figure}
\plotone{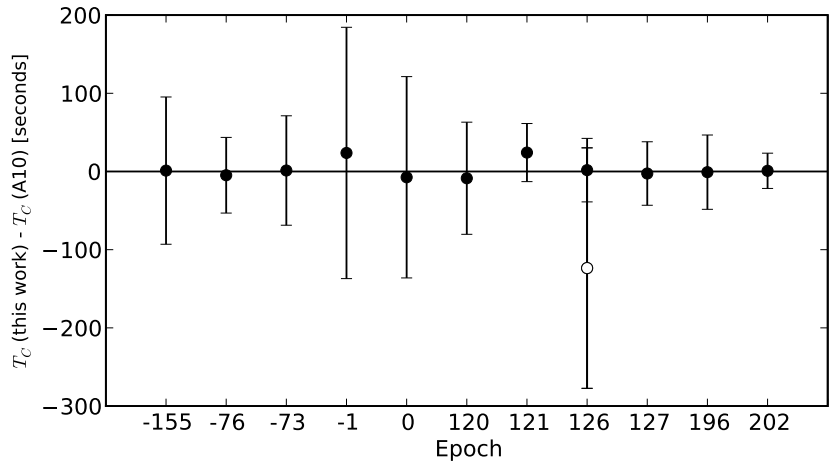}
\caption{Comparison of the eleven mid-transit times, \tc, obtained by A10 and recomputed by us in this work. Solid circles show the difference between the \tc ~measured by each group. The open circle shows the \tc ~for the new transit we observed in 2008-05-12 UT, and coincides with one of the transits measured by A10.  Although we find that the transit occurs \diffpoint ~seconds earlier, both results are consistent within the errors.
\label{comparacion}}
\end{figure}

\clearpage

\begin{figure}
\epsscale{.80}
\plotone{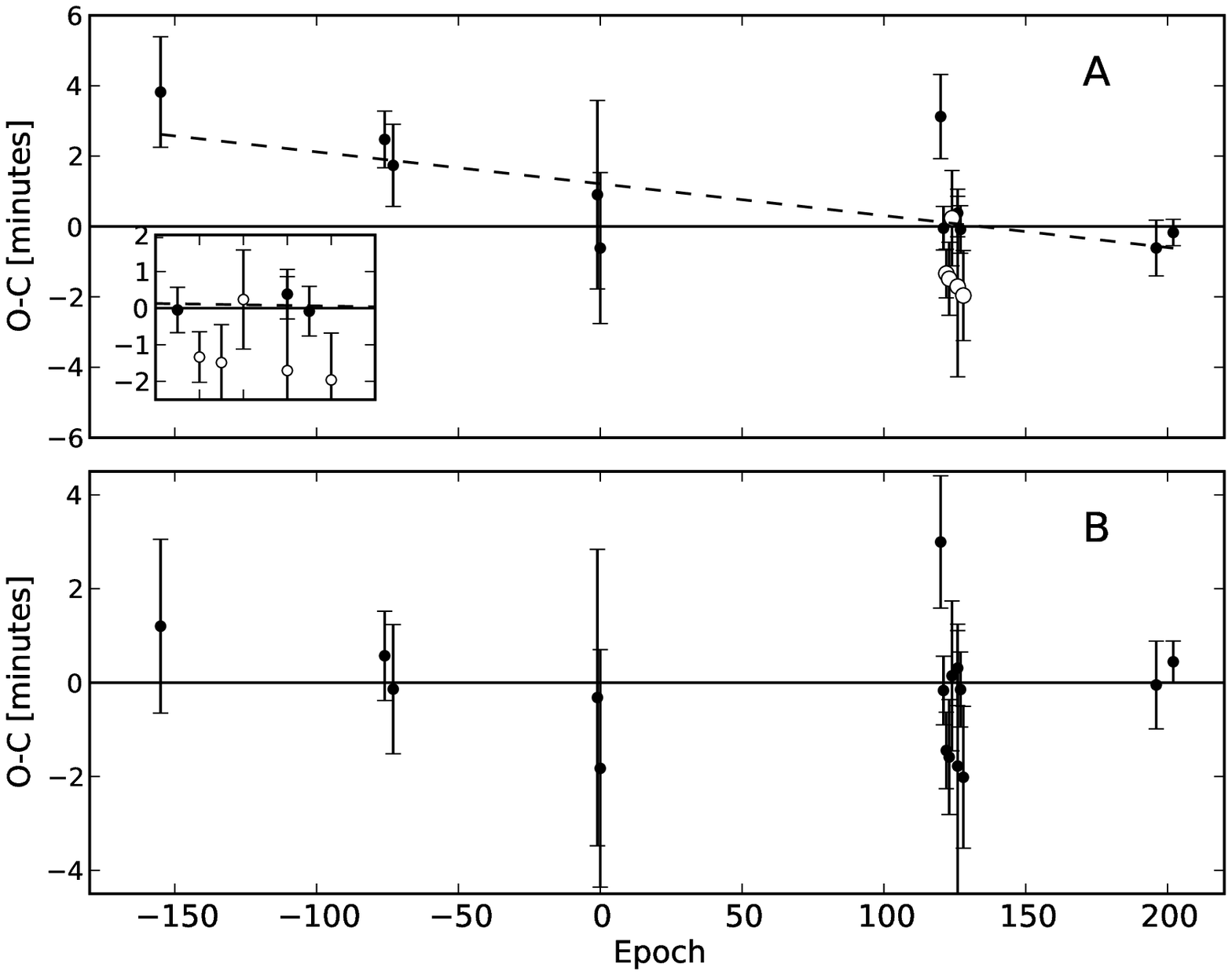}
\caption{\textbf{Panel-A:} \textit{Observed minus Calculated} diagram of the central times of the transits of OGLE-TR-111b. Black dots represent the times obtained with data from \cite{pawel10}, \cite{W07}, \cite{D08} and \cite{adams10a}, while the white dots are from the transits of this work.  The dashed line represents a linear fit of the data.  In the small box a zoom of the points between 110th and 120th epochs is shown. \textbf{Panel-B:} When the linear trend is removed no variations of more than \ttvamplitudelimit ~are present.  The errors were rescaled by $\sqrt{\chi^2}=1.18$ to make them consistent with a linear fit with $\chi^{2}_{red}=1$. \label{o-c}}
\end{figure}

\clearpage

\begin{figure}
\plotone{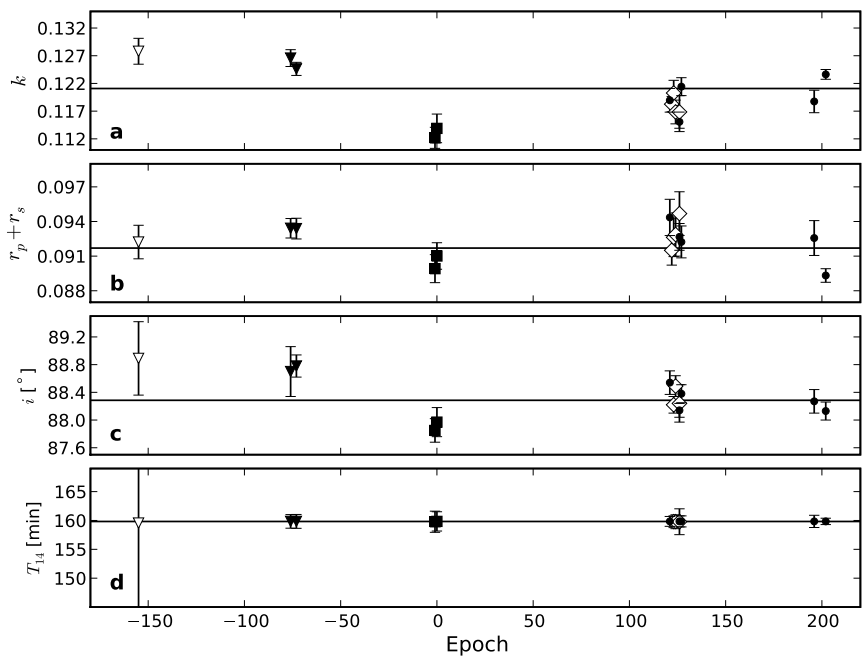}
\caption{ Resulting values for the ratio of the radii (\textbf{Panel a}), sum of the fractional radii (\textbf{Panel b}), orbit's inclination (\textbf{Panel c}) and duration of the transit (\textbf{Panel d}), defined as the time between the first and fourth contact, for each light curve using JKTEBOP. Open and solid triangles correspond to \cite{pawel10} and \cite{W07} transits.  Solid squares correspond to the two transits of \cite{D08} and solid dots represent the transits of \cite{adams10a}.  Open diamonds correspond to the new transits of this work. 
  \label{ki}}
\end{figure}

\clearpage

\begin{figure}
\plotone{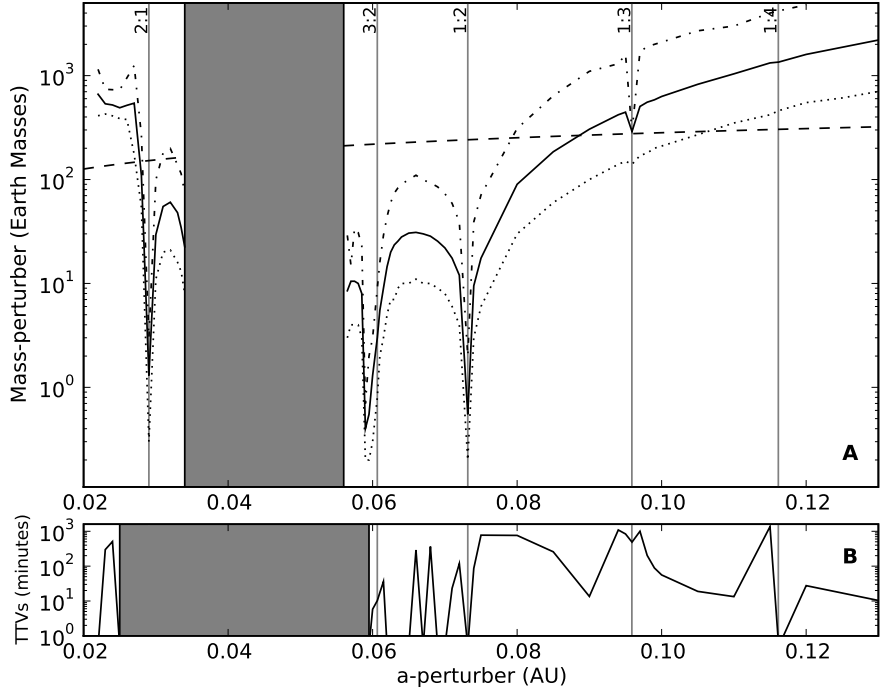}
\caption{\textbf{Panel-A:} Upper mass limits of an orbital perturber.  These simulations were computed using $e=0$. The solid line represents transit timing variations of 1.5 minutes.   The dotted line and dash-point line represent TTVs of 0.5 and 5 minutes, respectively.   The dashed line corresponds to the limits due the radial velocities observations. Vertical lines and gray strip indicate the orbital resonances locations and the instability region respectively. An orbital companion of OGLE-TR-111b should have a mass in the region below the black solid line which corresponds to the mass limit imposed by the timing analysis. 
\textbf{Panel-B:} Transit Timing Variations with $e=0.3$. If the eccentricity of the perturber represented by the solid line in Panel-A is increased, it will exhibit larger values than 1.5 minutes for its TTVs. Regions with TTVs below 1 minute correspond to unstable orbits with this new configuration.  \label{mvsa}}
\end{figure}

\end{document}